\documentclass[10pt]{iopart}
\expandafter\let\csname equation*\endcsname\relax
\expandafter\let\csname endequation*\endcsname\relax
\usepackage{todonotes}
\usepackage{amsmath}
\usepackage{psfrag}
\usepackage{graphicx}
\usepackage{graphics}
\usepackage{hyperref}
\usepackage{color}
\usepackage{siunitx}
\usepackage{verbatim}
\usepackage{physics}
\usepackage[acronym]{glossaries}
\usepackage[normalem]{ulem}
\usepackage{mathrsfs}
\usepackage{mathtools}
\usepackage{subcaption}
\newcommand{\beq}{\begin{equation}}
\newcommand{\eeq}{\end{equation}}
\newcommand{\beqa}{\begin{eqnarray}}
\newcommand{\eeqa}{\end{eqnarray}}
\newcommand{\ba}{\begin{aligned}[b]}
\newcommand{\ea}{\end{aligned}}

\usepackage[backend=biber, 
style=nature, 
doi=false,
isbn=false,
url=false,
eprint=false]{biblatex}
\defbibfilter{optica}{keyword={optica}}
\usepackage{placeins}

\usepackage{stfloats}
\usepackage{graphicx}
\graphicspath{{figures/}}
\usepackage{amsmath}
\usepackage{amssymb}
\usepackage{array}
\usepackage{physics}
\usepackage{multirow}
\usepackage{siunitx}
\usepackage[normalem]{ulem}
\usepackage[acronym]{glossaries}
\usepackage{todonotes}
\usepackage{subfig}
\usepackage{subfloat}
\usepackage{url}
\usepackage{stfloats}
\usepackage{mathrsfs}

\definecolor{darkgreen}{rgb}{0.0, 0.5, 0.0}
\definecolor{c2}{rgb}{0.3, 0.0, 0.6}
\definecolor{c3}{rgb}{0.6, 0.2, 0.1}

\newcommand{\cb}[1]{\color{black}#1\color{black}}

\DeclareSIUnit\baud{baud}
\DeclareSIUnit{\dBm}{dBm}
\DeclareSIUnit{\dB}{dB}

\DeclareGraphicsExtensions{.png, .pdf, .pgf}
\graphicspath{{../results/}{../immagini/}}
\makeatletter
\providecommand{\sf@counterlist}{}
\makeatother

\AtEveryBibitem{%
  \clearfield{note}%
  \clearfield{language}%
  \clearfield{issn}%
  \clearfield{month}%
  \clearfield{url}%
}

\begin{document}
\title[Variational treatment of an Optical Cavity with a Moving Mirror]{Variational Treatment of the  Electromagnetic Field in an Optical Cavity with a Moving Mirror}

\author{F. Lorenzi$^{1, *}$, 
M. G. Pelizzo$^{1,3,5}$, and L. Salasnich$^{2,4, 6 ,7}$}

\address{
$^{1}$Dipartimento di Ingegneria dell'Informazione, 
Universit\`a di Padova, via Gradenigo 6A, 35131 Padova, Italy\\
$^{2}$Dipartimento di Fisica e Astronomia ``Galileo Galilei", 
Universit\`a di Padova, Via Marzolo 8, 35131 Padova, Italy\\
$^{3}$Centro di Ateneo Studi e Attività Spaziali, Universit\`a di Padova, 
viale Venezia1 , 35131 Padova, Italy\\
$^{4}$Istituto Nazionale di Fisica Nucleare (INFN), Sezione di Padova, via Marzolo 8, 35131 Padova, Italy\\
$^{5}$Istituto di Fotonica e Nanotecnologie, Consiglio Nazionale delle Ricerche, 
via Trasea 7, 35131 Padova, Italy\\
$^{6}$Padua QTech Center, Universit\`a di Padova, 
via Gradenigo 6A, 35131 Padova, Italy\\
$^{7}$Istituto Nazionale di Ottica (INO) del Consiglio Nazionale delle Ricerche (CNR), via Nello Carrara 1, 50019 Sesto Fiorentino, Italy \\
$^*$ email: \texttt{francesco.lorenzi.2@phd.unipd.it}
}
\maketitle
\begin{abstract}
    Optical cavities with moving mirrors provide a versatile platform for exploring radiation–matter interactions and optically mediated mechanical effects, whose control has wide technological implications. However, capturing the coupled dynamics of the electromagnetic field and of the mirror within a consistent theoretical framework remains challenging.
    We analyze the problem of the coupling between classical electromagnetic fields in a cavity and a movable mirror, considering both nonrelativistic and relativistic regimes of motion. 
    Starting from the equations of motion for a mirror subject to a generic external potential, we provide a variational formulation of the mirror–radiation interaction. Within this framework, a single-mode variational approximation is introduced, which captures the essential dynamical features of the coupled system. In the special case of a mirror undergoing free motion, the variational method yields an exact solution. This unified treatment highlights the connection between different dynamical regimes and provides a basis for analyzing applications ranging from precision interferometry to relativistic radiation-pressure effects.
\end{abstract}

\section{Introduction}

The idea that light carries momentum and exerts pressure dates back to Kepler and Newton \cite{ashkin1972pressure}. Maxwell’s 1873 electromagnetic theory provided a mechanical framework, later confirmed experimentally in the early 20th century by Lebedev \cite{lebedev} and by Nichols and Hull \cite{nichols1901preliminary, nichols1903pressure}.
Today, radiation pressure is studied across vastly different length scales.  
At the microscopic level, the interaction between a quantized electromagnetic field and the quantized motion of matter acting as a mirror has attracted growing attention, motivated in part by the development of ultrasensitive interferometers for precision measurements. From the early predictions of light-induced ponderomotive effects by Braginskii, Manukin and Tikhonov \cite{braginski1967ponderomotive, braginskiiInvestigation1970}, the subject has matured into a field that bridges optics and mechanics at both the classical and quantum levels \cite{aspelmeyerCavityOptomechanics2014,kippenbergCavityOptomechanicsBackAction2008a,meystreShortWalkQuantum2013}.  
Initial studies identified parametric instabilities in Fabry–Perot interferometers \cite{braginskyParametricOscillatoryInstability2001} and proposed light-induced damping as a stabilization mechanism \cite{braginskyLowQuantumNoise2002}. Subsequent experiments with high-$Q$ microcavities confirmed radiation-pressure–induced oscillations \cite{kippenbergAnalysisRadiationPressureInduced2005} and revealed nonlinear multistability \cite{marquardtDynamicalMultistabilityInduced2006}.
Further theoretical developments on sideband cooling and dynamical backaction \cite{marquardtQuantumTheoryCavityAssisted2007,wilson-raeTheoryGroundState2007} opened new routes toward cooling mechanical motion to its quantum ground state \cite{groblacherDemonstrationUltracoldMicrooptomechanical2009}. More recently, advances in quantum state engineering and tomography \cite{meystreShortWalkQuantum2013, vannerCoolingbymeasurementMechanicalState2013, riedingerNonclassicalCorrelationsSingle2016, vijayanCavitymediatedLongrangeInteractions2024, verhagenQuantumcoherentCouplingMechanical2012}, wavelength conversion \cite{notomiWavelengthConversionDynamic2006,prebleSinglePhotonAdiabatic2012,jirauschekWavelengthShiftingIntracavity2015,kranendonkModelessOperationWavelengthagile2005}, and ultrafast tuning have established optomechanics as a powerful platform for both fundamental science and emerging technologies \cite{daniaHighpurityQuantumOptomechanics2025, huangRoomtemperatureQuantumOptomechanics2024}. A striking example is the development of extremely sensitive interferometers for gravitational-wave detection.
At the opposite extreme, on astronomical length scales, radiation pressure has been proposed as a means of spacecraft propulsion, enabling relativistic speeds for lightweight ``lightsails,'' i.e. reflective surfaces carrying a payload \cite{davoyan_photonic_2021, kulkarni2018relativistic, atwaterMaterialsChallengesStarshot2018f, ilicSelfstabilizingPhotonicLevitation2019e}. In this regime, the reflective surface is modeled as moving freely.  
Both microscopic and macroscopic scenarios can be treated within cavity-like configurations, with the reflective surface acting as a dynamical boundary \cite{bae2021photonic, bae2022photonic}.

Theoretical approaches strongly depend on the regime of interest. The quantum dynamics of a coupled mirror–radiation system has been analyzed via field mode decomposition and construction of the corresponding Hamiltonian operator \cite{lawEffectiveHamiltonianRadiation1994a,law1995interaction,Law2012, razavyQuantumRadiationOnedimensional1985}. This framework captures phenomena such as the dynamical Casimir effect, arising from nonadiabatic dynamics \cite{calucci1992casimir}. For relativistic motion, exact solutions exist only in the case of uniform mirror velocity, obtained through coordinate transformations \cite{baranov1967electromagnetic, moore1970quantum}, which yield quasi-modes.

In this article, we present an approximate method to compute the full dynamics of a mirror coupled to cavity radiation in the classical regime. Our approach employs a variational single-mode approximation, resulting in an effective potential for the mirror. Since this treatment remains valid in the relativistic regime, it also applies to the motion of a free lightsail under laser radiation pressure. 
The article is organized as follows. In Section~2 we review the fundamental equations for the electromagnetic field and the mirror, and we perform the modal decomposition of the cavity field.
In Section~3 we discuss how the dynamics can be analyzed in terms of a variational single-mode approach, introduce the appropriate variational ansatz, and derive the equations of motion in the relativistic case, including the case of a generic external potential. Section~4 discusses how to solve the resulting equations by means of a simplification and the Wentzel-Kramers-Brillouin (WKB) method, and obtain the resulting dynamics. The appendix is dedicated to the special case of relativistic motion in presence of a harmonic potential, which is interesting from the mathematical point of view.

\section{Fundamental Theory}
The radiation pressure effect can be formalized by imposing appropriate boundary conditions and calculating the resulting force on the mirror’s surface. We consider a \cb{ideally } reflecting mirror, modeled as a flat perfect conductor. In this case, the boundary condition requires the tangential component of the electric field to vanish at the mirror surface. However, when the mirror is in motion, this condition must be modified: due to Lorentz transformations, the electric and magnetic fields are not invariant between the laboratory and co-moving frames. The corresponding transformations for the electromagnetic fields are given by
\begin{equation}
\begin{aligned}
& \mathbf{E}^{\prime}=\gamma \ (\mathbf{E}+c\boldsymbol{\beta} \times \mathbf{B})-\frac{\gamma^2}{\gamma+1} \boldsymbol{\beta}(\boldsymbol{\beta} \cdot \mathbf{E}) \\
& \mathbf{B}^{\prime}=\gamma\left(\mathbf{B}-\frac{\boldsymbol{\beta}}{c^2}\times \mathbf{E}\right)-\frac{\gamma^2}{\gamma+1} \boldsymbol{\beta}(\boldsymbol{\beta} \cdot \mathbf{B}),
\end{aligned}
\end{equation}
where $\mathbf{E}^{\prime}$ and $\mathbf{B}^{\prime}$ are the electric field and magnetic induction fields in the frame moving with the mirror, $\mathbf{E}$ and $\mathbf{B}$ the ones in the lab reference frame, ${\boldsymbol \beta}={\bf v}/c$ is the relative velocity ${\bf v}$ in units of the speed $c$ of light in vacuum, and $\gamma=1/\sqrt{1-\beta^2}$ is the Lorentz-FitzGerald factor.
The fields are assumed to be transverse to the direction of movement of the mirror, that occurs along the $x$ direction.
Let the coordinate of the moving mirror be $q(t)$, such that $q(0)=q_0$ is its position at time $t=0$. The fixed mirror is instead located at the position $x=0$.
The boundary condition of vanishing transverse electric field on the mirror is imposed as
\begin{equation}
\mathbf{E}(q(t), t)+ c \boldsymbol{\beta} \times \mathbf{B}(q(t), t) = 0.
\end{equation}
In a one-dimensional cavity with propagation in vacuum, we assume a transverse electromagnetic field and, without loss of generality, retain only a single polarization, so that the fields can be written as $\mathbf{E} = E \ \mathbf{u}_y$ and $\mathbf{B} = B\ \mathbf{u}_z$.
Considering a motion of the mirror in the $\hat{x}$ direction, $\mathbf{v} = \dot{q}(t) \ \mathbf{u}_x$, the equations reduce to
\begin{equation}
\begin{aligned}
& E^{\prime}=\gamma(E+ \dot{q}(t) \ B),\
& B^{\prime}=\gamma\left(B-\frac{\dot{q}(t)}{c^2}E\right),
\end{aligned}
\end{equation}
with the boundary condition
\begin{equation}\label{eq:boundary}
E(q(t), t) - \dot{q}(t) \ B(q(t), t) = 0.
\end{equation}
This equation, together with $E(0, t)=0$, constitutes the time-varying boundary condition of the problem.
\cb{When the mirror is not ideally reflecting, part of the electromagnetic wave penetrates and is dissipated inside the material. The boundary conditions must then be changed since the tangential electric field is no longer null at the mirror surface, but it is related to the induced surface current and material response which can be modeled through a surface impedance \cite{someda}. A finite surface impedance accounts for absorption and phase shift upon reflection. }
The magnetic vector potential $\mathbf{A}$ and the electric scalar potential $\phi$ are introduced in the Coulomb gauge. In the absence of charges and currents in the domain of interest, the Coulomb gauge fixing corresponds to $\nabla^2\phi = 0$ and $\nabla \cdot \mathbf{A} = 0$. In this gauge, with $\mathbf{A} = A \ \mathbf{u}_y$ the fields are expressed as
\begin{equation}
\begin{aligned}
& E=-\pdv{A}{t} \,, \
& B=\pdv{A}{x} \,.
\end{aligned}
\end{equation}
The boundary condition (\ref{eq:boundary}) is implied by the requirement that the magnetic potential is constant, since the total derivative
\begin{equation}
{dA(q(t), t)\over dt} = -E(q(t), t) + \dot{q}(t) \ B(q(t), t) = 0
\end{equation}
corresponds to the boundary condition. The resulting boundary conditions are \cite{moore1970quantum}
\begin{align} 
A(0, t) = 0 \,, \label{eq:boundary1} \\ 
A(q(t), t)=0 \,.\label{eq:boundary2} 
\end{align}
\begin{figure}
\centering
\includegraphics[width=0.8\linewidth]{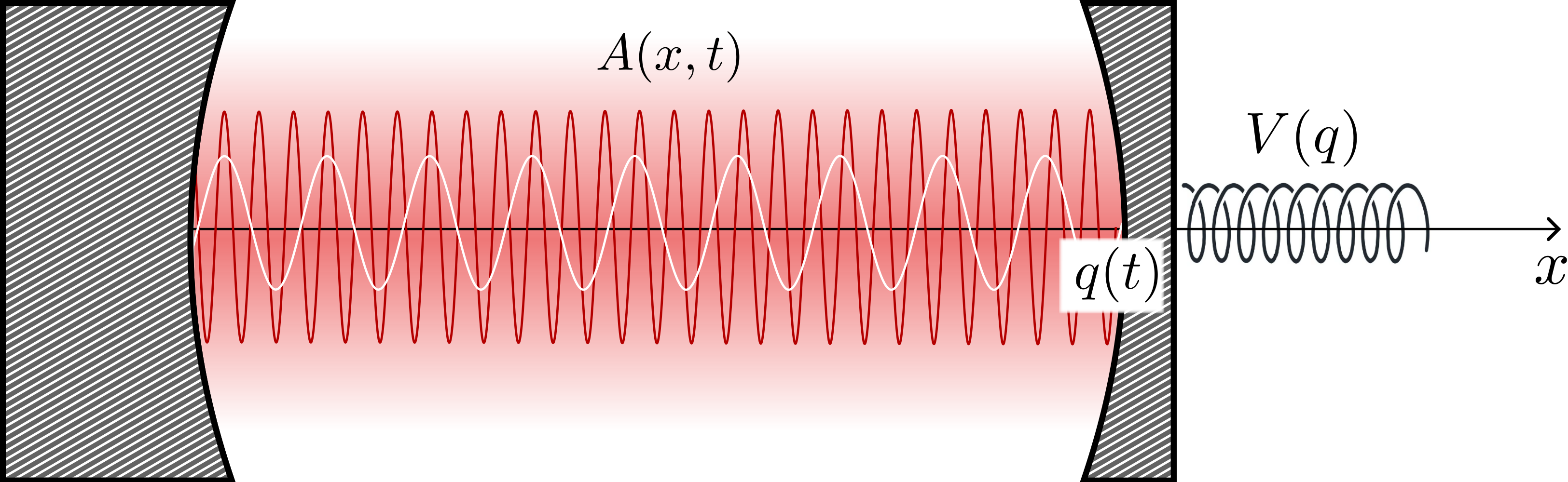}
\caption{Schematic representation of the cavity system. The fixed mirror is on the left, at position $q(t)$, and the movable mirror is on the right. Inside the cavity, the field $A(x, t)$, i.e. the transverse component of the magnetic vector potential, can have a number of excited modes, represented pictorially with sinusoids. An external potential $V(q)$ can be set to act on the moving mirror.}
\label{fig:schematic}
\end{figure}
Inside the cavity, in the absence of a medium, it holds
\begin{equation}\label{eq:Awaves}
\pdv[2]{A}{x}- \frac{1}{c^2}\pdv[2]{A}{t} = 0.
\end{equation}
The relativistic Newton's law for the mirror is given by the computation of the pressure element of the electromagnetic stress tensor in the laboratory reference frame. 
\cb{
If the mirror is not ideally reflective, the radiation pressure must balance not only the impinging and reflected momentum density, but also the effects of transmission and energy absorption into the material \cite{lorenzi-prapplied}. }
Assume the surface of the mirror is $S$. Since the shear terms are null on the mirror surface, only a force component in the $x$ direction, collinear with the mirror velocity vector, is expected. Such force, $\mathbf{F}^{(\mathrm{rad})} = F_x^{(\mathrm{rad})} \mathbf{u}_x$, is expressed as
\begin{equation}
F_x^{(rad)} = \frac{S}{2\mu_0} \left(\pdv{A}{x} + \frac{\dot{q}}{c^2} \ \pdv{A}{t}\right)^2\Bigg|_{x = q(t)} .
\label{itisdifferent}
\end{equation}
The total force is composed of $F_x^{\mathrm{rad}}$ and the force due to an external potential where the mirror is located. The corresponding force is assumed to act only along the axis of the cavity, with a magnitude of $-\partial V/\partial q$, hence neglecting transverse motion. The system is represented schematically in Fig.~\ref{fig:schematic}.
Taking into account the relativistic motion of the moving mirror, with momentum $\gamma m {\dot q}$ with $\gamma=1/\sqrt{1-{\dot q}^2/c^2}$ the Lorentz-FitzGerald factor, the relativistic Newton law gives (see also \cite[Appendix A]{lorenzi-prapplied})
\begin{equation}
\label{eq:relativistic_Newton}
\gamma^3 m \ddot{q} = F_x^{(\mathrm{rad})} -\frac{\partial V}{\partial q} \; .
\end{equation}
The nonrelativistic equation of motion is retrieved in the limit case of $\gamma \sim 1$. The boundary condition for the electromagnetic field instead needs to retain the same form of Eq.~(\ref{eq:boundary2}).

Following Law \cite{law1995interaction}, the electromagnetic field can be decomposed in a set of orthonormal functions satisfying the boundary conditions, as
\begin{equation}\label{eq:iiii}
A(x, t) = \sum_{\ell=1}^\infty Q_\ell(t)\sqrt{\frac{2}{q(t)}}\sin\left(\frac{\pi\ell}{q(t)} x\right) \,.
\end{equation}
Such decomposition leads to an expression of coupled-mode equations for Eq.~(\ref{eq:Awaves}) for the field, and Eqs.~(\ref{itisdifferent}, \ref{eq:relativistic_Newton}) for the mirror position:
\begin{equation}\label{eq:Qell}
\ddot{Q}_\ell = -\omega_\ell^2 Q_\ell + \frac{2\dot{q}}{q} \sum_j g_{\ell j}\dot{Q}_j + \frac{\ddot{q}q - \dot{q}^2}{q^2} \sum_j g_{\ell j}Q_j + \frac{\dot{q}^2}{q^2} \sum_{jh} g_{j\ell}g_{jh}Q_h,
\end{equation}
\begin{equation}\label{eq:q}
\gamma^3m\ddot{q} = -\frac{\partial V}{\partial q} + \frac{1}{q} \sum_{\ell,j} (-1)^{\ell+j}\omega_\ell \omega_j Q_\ell Q_j.
\end{equation}
with
\begin{equation}\label{eq:gkj}
g_{kj}=\begin{cases}
(-1)^{k+j}\dfrac{2kj}{j^2-k^2}, & k\neq j,\\
0, & k=j.
\end{cases}
\end{equation}
and $\omega_\ell = c\ell\pi/q$.
These equations are difficult to solve directly due to the all-to-all coupling of the modal amplitude equations.

The set of equations obtained constitutes the relativistic generalization of the dynamical boundary-value problem originally addressed by Razavy \cite{razavyQuantumRadiationOnedimensional1985} and by Law \cite{lawEffectiveHamiltonianRadiation1994a,law1995interaction}. In the nonrelativistic regime, one can recover Law’s effective Hamiltonian description directly by taking the appropriate limit, which corresponds to removing the factor $\gamma^{3}$ from Eq.~(\ref{eq:q}). In both relativistic and nonrelativistic cases, unlike the static situation, the frequency of each cavity mode is no longer constant: not only does the wavenumber depend on the mirror position, but the mirror’s motion also couples all modes to one another.

The presence of the coupling coefficients $g_{\ell j}$ indicates that every field mode interacts with all the others, with the mirror mediating energy exchange across the entire spectrum. This multimode interaction generally prevents closed-form solutions. The complexity is somewhat reduced when the mirror moves with uniform velocity, but even then a full multimode solution remains out of reach. Direct solution of the complete set of equations is therefore intractable, and approximations are essential.

%
%
\section{Single Mode Variational Treatment}
In the following, we construct a single-mode truncation using a variational approach, and subsequently apply an adiabatic approximation to the single-mode field. This framework provides a systematic reduction of complexity while retaining the essential physics of the mirror–field interaction.

The general Lagrangian structure of the problem is reviewed in the following. 
The Lagrangian of the system under investigation is given by 
\begin{equation} 
L = L_q + L_A \; , 
\end{equation}
where
\begin{equation}
L_q = - m c^2 \sqrt{1-{{\dot q}^2\over c^2}} - V(q) \,,
\end{equation}
is the relativistic Lagrangian of the moving mirror with mass $m$ and position $q(t)$, under an external potential $V(q)$. In the nonrelativistic limit, and neglecting constant terms, $L_q$ assumes the well-known form $L_q \approx m\dot{q}^2/2  - V(q)$.
The term associated to the radiation is instead
\begin{equation}\label{eq:marian}
L_A = {\epsilon_0 \, S \over 2}\int_0^{q} dx \left[ \left( {\partial A\over \partial t} \right)^2 - c^2 \left( {\partial A\over \partial x} \right)^2 \right] 
\end{equation}
is the Lagrangian of the electromagnetic field with 
vector potential $A(x,t)$ with $S$ the transverse area of the two mirrors. The two Lagrangians are coupled due to the upper limit of integration in $L_A$.  

The variational ansatz can be developed starting from the assumption that only one of the modes is excited, and mode coupling is negligible. 
Selecting only one mode from Eq.~(\ref{eq:iiii}), we obtain the following ansatz for the field $A(x,t)$
\begin{equation}\label{eq:varian}
A(x, t)= Q(t) \sqrt{\frac{2}{q(t)}} \sin \left({\pi \, \ell \over q(t)} x\right) , 
\end{equation}
This is the very special case of a monochromatic electromagnetic wave, where only one single mode, with the mode number $\ell$, wavenumber $k_\ell(t)=\pi \ell/q(t)$ and amplitude $Q(t)$, is occupied. This single-mode ansatz  satisfies the required boundary conditions Eqs.~(\ref{eq:boundary1}, \ref{eq:boundary2}),
and $Q(t)$ is the time-dependent variational parameter.
\cb{
We note that the ansatz of Eq.~\eqref{eq:varian} bears some similarity to the standard adiabatic mode-elimination description used in cavity optomechanics: a single electromagnetic cavity mode, whose instantaneous frequency $\omega_\ell(t)=\pi c\,\ell/q(t)$ is parametrically modulated by the mirror position $q(t)$~\cite{jiang2016dynamics}, while the other modes are effectively accounted for through averages, assuming they oscillate much faster. In the quantum limit, this reduces to the single-mode optomechanical model, in which the field couples to the mechanical displacement~\cite{ghorbani2025effects}. Equation~\eqref{eq:varian}, however, represents a stronger approximation, as it corresponds to setting all modes except the one under consideration to zero amplitude in Eqs.~\eqref{eq:Qell} and~\eqref{eq:q}, and neglecting their mutual coupling.

The validity of the single-mode truncation can be assessed quantitatively by estimating the effect of velocity-induced intermode couplings. Using the previously discussed mode-coupling coefficients~\eqref{eq:gkj}, the evolution equation~\eqref{eq:Qell} for mode $\ell$ contains mixing terms that depend on the amplitudes and derivatives of the amplitudes of other modes. Since in the following we consider large mode numbers $\ell$, we examine the asymptotic behaviour of the coupling coefficients as $\ell$ becomes large. The dominant contribution in this regime corresponds to the last coupling term in Eq.~\eqref{eq:Qell}, involving the neighbouring modes with $j=\ell\pm1$. Indeed, for large $\ell$, the coupling coefficient scales as $|g_{\ell,\ell\pm1}|\sim \ell$. Since $\omega_\ell = \pi c\,\ell/q$, the dimensionless ratio controlling the magnitude of these cross terms relative to the optical frequency in the last contribution is
\begin{equation}
\frac{(\dot q/q)\,|g_{\ell,\ell\pm1}|}{\omega_\ell}
\sim
\frac{(\dot q/q)\,\ell}{\pi c\,\ell/q}
=
\frac{|\dot q|}{\pi c},
\end{equation}
which is independent of $\ell$.
More distant modes are even less relevant: for $j=\ell+m$ with $m>1$, one finds $|g_{\ell,\ell+m}|\sim \ell/m$, so the corresponding terms are further suppressed by the factor $1/m$. All other terms vanish asymptotically as $\ell$ becomes large. Therefore, we conclude that the single-mode ansatz is reliable in the nonrelativistic regime, but many mode interactions terms are suppressed in the case of large $\ell$. 
 }
Remarkably, this ansatz is treatable analytically to a very good degree. Inserting Eq.~(\ref{eq:varian}) into Eq.~(\ref{eq:marian}) and integrating over $x$ we obtain 
\begin{equation}\label{discola}
L_A = {\epsilon_0 S\over 2} \left[
\left( {\dot Q}^2 - {c^2\pi^2\ell^2\over q^2} Q^2\right) 
+ 
\left(1+{2\pi^2 \ell^2\over 3}\right)
{{\dot q}^2Q^2\over q^2} \right]
\end{equation}
As usual, the conjugate momenta of $q(t)$ and $Q(t)$ are defined as 
\begin{eqnarray} 
p_q &=& {\partial L\over \partial {\dot q}} = {m {\dot q}\over \sqrt{1-{{\dot q}^2\over c^2}}}\,,
\\
p_Q &=& {\partial L\over\partial {\dot Q}} = \epsilon_0 S {\dot Q} \,,
\end{eqnarray}
while the Hamiltonian 
\begin{equation} 
H = p_q {\dot q} + p_Q {\dot Q} - L 
\end{equation}
reads 
\begin{equation} 
H = H_q + H_Q + H_{\mathrm{cross}} \,,
\label{ham}
\end{equation}
with the energy associated to the mirror
\begin{equation}\label{eq:Hq}
    H_q = \frac{mc^2}{\sqrt{1-\frac{\dot{q}^2}{c^2}} }+ V(q) \, , 
\end{equation}
and the energy associated to the mode,
\begin{equation}
    H_Q = \frac{\epsilon_0 S}{2}\left(\dot{Q}^2+\frac{c^2\pi^2\ell^2}{q^2}Q^2\right) \, ,
\end{equation}
and the cross-energy term,
\begin{equation}\label{eq:h-discolo}
    H_{\mathrm{cross}} = \frac{\epsilon_0 S}{2} \left(1+\frac{2\pi^2\ell^2}{3} \right) \ \frac{\dot{q}^2Q^2}{q^2} \, .
\end{equation}
This Hamiltonian is a conserved quantity, in particular, it is the total energy of the effective model of our system.
\cb{The canonical variables of our single-mode treatment can be associated to the ones utilized in the foundational Hamiltonian framework of Law~\cite{law1995interaction} as follows. Once a fixed mode index $\ell$ is chosen, the dynamical electromagnetic mode amplitude $Q$ of the present theory is associated with $Q_\ell$ of the foundational model (see also Eq.~\eqref{eq:Qell}), and the mirror coordinate $q$ is retained without modification. This identification, and the corresponding relations for the canonical momenta, becomes exact in the stationary-mirror limit, where mode mixing is absent, as described above. }
We get the following Euler-Lagrange equations.
The \cb{equation of motion } for $Q$ is simple: 
\begin{equation}
{\ddot Q} + \left[{c^2\pi^2 \ell^2\over q^2} - \left(1+\frac{2\pi^2\ell^2}{3}\right) \frac{\dot{q}^2}{q^2} \right] Q = 0  
\end{equation}
and the one for $q$ is a bit less simple
\begin{align}
    &{m \, \ddot{q}\over \left(1-{{\dot q}^2\over c^2}\right)^{3/2}} + \frac{\partial V}{\partial q} -  \epsilon_0 S \Bigg[\frac{c^2\pi^2\ell^2}{q^3}Q^2 \nonumber \\ &- \left(1+\frac{2\pi^2\ell^2}{3}\right) \left(-\dot{q}^2\frac{Q}{q} + 2\dot{q}\dot{Q}+\ddot{q}Q\right)\frac{Q}{q^2}\Bigg]  = 0
\end{align}
The above equations represent a major simplification of the original coupled-modes ones Eqs.~(\ref{eq:q}-\ref{eq:Qell}). They can be further simplified as detailed in the following section, where we apply the WKB method to a reduced version of the Lagrangian.
We demonstrate that, although in general only a numerical approach is viable, in the case of a free mirror it is possible to obtain an analytical solution.
\section{Solution Method}
We observe that the variational Lagrangian (\ref{discola}), contains a coupling term between ${\dot q}(t)$ and $Q(t)$, which can be neglected if it is the case that $\dot{q}\ll c$ (see also Ref. \cite{Law2012}).
\cb{It is worth noting that this corresponds to a hybrid approximation scheme, in which we require the velocity not to be ultrarelativistic in order to neglect a cross-energy term in Eq.~\eqref{eq:h-discolo}, but we keep the relativistic form of the kinetic energy of the mirror. This aspect will be stressed in the case of relativistic solutions in Section~\ref{sec:libero}. }
By applying this approximation, we have the modified effective Lagrangian
\begin{equation}\label{eq:simplicio}
L = - m c^2 \sqrt{1-{{\dot q}^2\over c^2}} + 
{\epsilon_0 S\over 2} 
\left( {\dot Q}^2 - {c^2\pi^2\ell^2\over q^2} Q^2\right) \; . 
\end{equation}
The corresponding Euler-Lagrange equations are 
\begin{eqnarray}
\frac{m\ddot{q}}{\left(1-\frac{\dot{q}^2}{c^2}\right)^{3/2}}- {\epsilon_0 S} {c^2\pi^2\ell^2\over q^3} Q^2 + \frac{\partial V}{\partial q} &=& 0 \label{q-simplicio}
\\
{\ddot Q} + {c^2\pi^2\ell^2\over q^2} Q &=& 0  \label{eq:Qslow}
\end{eqnarray}

%
%
The characteristic scale of $\ell$ is determined by comparing the light wavelength and the system length. We recall the definition
\begin{equation}\label{eq:ell}
    \ell = \frac{2 q_0}{\lambda_0} = \frac{q_0\omega_0}{2\pi c} \, ,
\end{equation}
where $q_0=q(0)$, and $\omega_0 = \omega_0(q_0)$.
Consider for example a typical infrared laser, with a wavelength of $\lambda_0\sim$\SI{1}{\micro m}, and an initial distance corresponding to $q_0 \sim$ \SI{1}{mm}. In this case the mode index is $\ell\sim 10^{3}$. 
\cb{In these conditions, if we consider a mirror in a harmonic potential, the optical oscillation period $T_{\mathrm{opt}} = 2\pi/\omega_0 \sim  10^{-15}\,\mathrm{s}$ is many orders of magnitude shorter than the typical experimentally achievable mechanical period $T_{\mathrm{m}} = 2\pi/\Omega$, which for $\Omega \sim 2\pi\times 10^{7}\,\mathrm{s^{-1}}$ \cite{aspelmeyerCavityOptomechanics2014} gives $T_{\mathrm{m}} \sim 10^{-7} \,\mathrm{s}$ , thus yielding a timescale separation of approximately $T_{\mathrm{m}} / T_{\mathrm{opt}} \sim 10^{8}$. For the case of freely moving mirror, the relevant time scale of the mechanical motion is the value $T'_\mathrm{m}=q(t)/|\dot{q}(t)|$, which, compared to the instantaneous period associated to the optical oscillations $c\pi\ell/q$ gives the slowly varying condition as $|\dot{q}|/(c\pi\ell)\ll 1$. For a sufficiently high value of $\ell$, this can be considered to hold for every time instant. }
This suggest the usage of short-wavelength asymptotics, by means of the WKB method. The WKB approach provides an approximate solution of Eq.~(\ref{eq:Qslow}), namely a solution in which the mode amplitude $Q(t)$ is provided as a function of time and $q(t)$.
The first-order (eikonal and transport) WKB for the mode amplitude, with initial conditions $Q(0)=Q_0$ and $\dot{Q}(0)=0$, is given by
\begin{equation}\label{eq:Q-wkb}
    Q(t) \approx Q_0 \sqrt{\frac{q(t)}{q_0}} \ \cos\left(\int_0^{t}d\tau \ \frac{c\pi\ell}{q(\tau)}\right) =  Q_0 \sqrt{\frac{q(t)}{q_0}} \ \cos\left(\int_0^{t}d\tau \ \omega_0 \frac{q_0}{q(\tau)}\right).
\end{equation}
By substitution into Eq.~(\ref{q-simplicio}), we obtain an integro-differential equation for $q$. The term $Q^2$ contains sum and difference frequencies as well as a constant component. 
We can neglect the fast oscillating terms, obtaining a simple differential equation for $q(t)$.
By substituting this solution into the equation for the position of the mirror (\ref{q-simplicio}), we get
\begin{equation}\label{eq:onlyq}
    \ddot{q} = \frac{(1-\frac{\dot{q}^2}{c^2})^{3/2}}{m}\left( \frac{\epsilon_0 S}{2} \ \frac{c^2 \pi^2 \ell^2}{q^2}  \ \frac{Q_0^2}{q_0} - \frac{\partial V}{\partial q}\right),
\end{equation}
that can be solved numerically.  

As an initial condition, we will suppose that a certain electromagnetic energy is stored in the mode  at $t=0$, and that $q(t) = q_0$ for $t\leq 0$. The corresponding energy is
\begin{equation}\label{eq:initial_energy}
E_0 = \frac{\epsilon_0 S}{2} Q_0^2 \omega_0^2 \, ,
\end{equation}
In the following, we will consider an harmonic potential of shape $V(q) =\Omega^2 m (q-\bar{q})^2/2$.
By assuming a time scale of $t_{\mathrm{rel}} = q_0/c$ and a length scale of $l_\mathrm{rel} = q_0$, defining $\mathfrak{q}=q/q_0$ and $\mathfrak{\bar{q}} = \bar{q}/q_0$, and using Eq.~(\ref{eq:onlyq}, \ref{eq:initial_energy}) we have the following normalized equation
\begin{equation}\label{eq:second-order-ode}
    \ddot{\mathfrak{q}} = \left(1-\dot{\mathfrak{q}}^2\right)^{3/2} \left(\frac{\Gamma}{\mathfrak{q}^2} - K (\mathfrak{q}-\bar{\mathfrak{q}})\right) \, .
\end{equation}
with $\Gamma = E_0/(mc^2)$, and $K=(\Omega q_0/c)^2$.

The equation of motion (\ref{eq:second-order-ode}) can be manipulated in the form of an implicit integral equation for the time variable in the following way. Substituting the time derivative with a derivative with respect to $\mathfrak{q}$ in $\ddot{\mathfrak{q}}$, we get
\begin{equation}
\dot{\mathfrak{q}}\dv{\dot{\mathfrak{q}}}{\mathfrak{q}}  = (1-\dot{\mathfrak{q}}^2)^{3/2}
\left(\frac{\Gamma}{\mathfrak{q}^2} -K(\mathfrak{q}-\bar{\mathfrak{q}})\right) \, ,
\end{equation}
therefore we can separate the equation
\begin{equation}
    \frac{\dot{\mathfrak{q}}}{(1-\dot{\mathfrak{q}}^2)^{3/2}}\dv{\dot{\mathfrak{q}}}{\mathfrak{q}} = \frac{\Gamma}{8 \mathfrak{q}^2} - K(\mathfrak{q}-\bar{\mathfrak{q}})\, .
\end{equation}
Integrating we get the energy conservation equation (see also Eq.~(\ref{eq:Hq}))
\begin{equation}
    \frac{1}{\sqrt{1-\dot{\mathfrak{q}}^2}} + \mathcal{V}(q) =\mathcal{E} \, .
\end{equation}
where the first two terms on the left hand side constitute the kinetic energy, $\mathcal{E}$ is the conserved energy, and $\mathcal{V}(\mathfrak{q})$ is an effective potential, defined as
\begin{equation}\label{eq:potential}
    \mathcal{V}(\mathfrak{q}) = \mathcal{V}_\Gamma(\mathfrak{q}) + \mathcal{V}_K(\mathfrak{q}) \,;
\end{equation}
separated in radiation pressure component
\begin{equation}\label{eq:VGamma}
    \mathcal{V}_\Gamma(\mathfrak{q}) = \frac{\Gamma}{\mathfrak{q}} \,,
\end{equation}
and elastic component
\begin{equation}\label{eq:VK}
    \mathcal{V}_K(\mathfrak{q}) = \frac{K(\mathfrak{q}-\bar{\mathfrak{q}})^2}{2} \,.
\end{equation}
By applying the initial conditions ($\mathfrak{q}(0)=1 \, , \dot{\mathfrak{q}}(0)=0$), we obtain the initial energy $\mathcal{E} = \Gamma + K(1-\bar{\mathfrak{q}})^2/2 +1$. After solving for the velocity $\dot{\mathfrak{q}}$ we obtain the implicit relation
\begin{equation}
    \dot{\mathfrak{q}} = \pm \sqrt{1-\frac{1}{(\mathcal{E}-\mathcal{V}(\mathfrak{q}))^2}}
\end{equation}
implying
\begin{equation}\label{eq:integralone}
    t = \pm \int_1^\mathfrak{q} \frac{ds \ |\mathcal{E}-\mathcal{V}(s)|}{\sqrt{(\mathcal{E}-\mathcal{V}(s))^2-1}} \,.
\end{equation}
This integral does not have a solution in terms of elementary functions in general.
It is equivalent to directly solve the second-order differential equation (\ref{eq:second-order-ode}). We notice that the choice of the numerical solver is critical to have the energy conservation, so we use a symplectic integration method of the Størmer–Verlet type \cite{geometric-numerical}.
We proceed to analyze the different limit cases. In the relativistic regime, a couple of special cases are important: the case of a pure relativistic harmonic oscillator ($\Gamma=0$), and the case of a mirror free to move ($K=0$). The case of relativistic motion in a harmonic potential (with $K\neq0$) is of mathematical interest, and it is treated in the Appendix A. The second case is instead interesting in the context of laser propulsion, and will be analyzed below.
We focus now on the nonrelativistic regime.

\subsection{Nonrelativistic motion in the harmonic potential}\label{sec:molla}
In the nonrelativistic regime, corresponding to most experimental setups, intuition on the general dynamics can be gained by studying the fixed point, its stability, and the corresponding small oscillations. 
The nonrelativistic Lagrangian can be obtained expanding the kinetic term in Eq.~(\ref{eq:simplicio}) in powers of $\frac{\dot{q}}{c}$ to the quadratic order. We also neglect the constant term corresponding to the relativistic rest energy, obtaining
\begin{equation}\label{eq:simplicio-NR}
L = \frac{m}{2} {\dot q}^2 + 
{\epsilon_0 S\over 2} 
\left( {\dot Q}^2 - {c^2\pi^2\ell^2\over q^2} Q^2\right) \; . 
\end{equation}
Following the WKB procedure for the elimination of the field $Q$ discussed above, the resulting nonrelativistic effective equation for the motion of the mirror is 
\begin{equation}\label{eq:onlyq-nr}
    \ddot{q} =\frac{\epsilon_0 S}{m} \ \frac{c^2 \pi^2 \ell^2}{q^2}  \ \frac{Q_0^2}{2q_0} - \frac{1}{m}\frac{\partial V}{\partial q},
\end{equation}
and in normalized units
\begin{equation}\label{eq:second-order-ode-nr}
    \ddot{\mathfrak{q}} = \frac{\Gamma}{\mathfrak{q}^2} - K (\mathfrak{q}-\bar{\mathfrak{q}})  \, .
\end{equation}
Therefore the motion of the mirror is just the motion of a nonrelativistic particle in the potentials $\mathcal{V}_{\Gamma}$ and $\mathcal{V}_K$ obtained in Eqs.(\ref{eq:VGamma}-\ref{eq:VK}).

\begin{figure}
  \centering
  \begin{subfigure}{0.54\textwidth}
    \includegraphics[width=\linewidth]{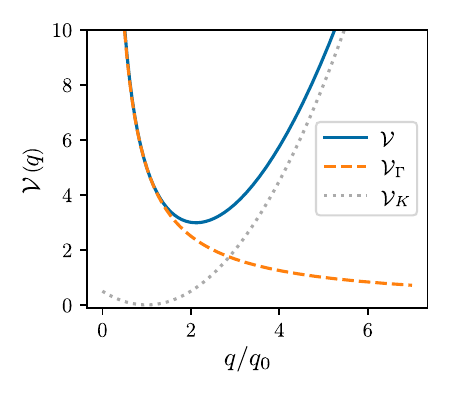}
    \caption{}
    \label{fig:1a}
  \end{subfigure}
  \hfill
  \begin{subfigure}{0.45\textwidth}
    \includegraphics[width=\linewidth]{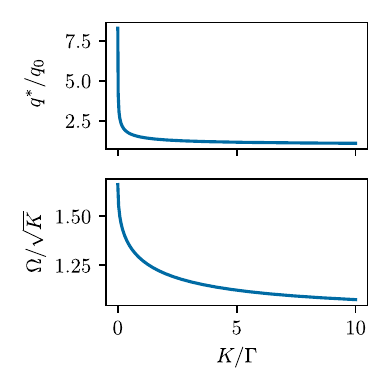}
    \caption{}
    \label{fig:1b}
  \end{subfigure}
  \caption{Effective potential and fixed-point properties of the dynamics of the mirror under the pressure of radiation approximated using the WKB approach. Panel (a) represents the effective potential $\mathcal{V}$ as a function of the coordinate $\mathfrak{q}$, showing the composition of the radiation pressure term $\mathcal{V}_\Gamma$ and the elastic term $\mathcal{V}_K$. The parameters are $K=1$ and $\Gamma=5$. Panel (b) represent the fixed point coordinate $q^*$ (top) and the frequency of small oscillations $\Omega$ (bottom), as a function of the ratio $K/\Gamma$.}
  \label{fig:fixed_point}
\end{figure}
To obtain the stationary solution $\mathfrak{q}^*$\cb{$=q^*/q_0$ } of the differential equation, we impose $\Gamma / \mathfrak{q}^{*2} - K(\mathfrak{q}^*-\bar{\mathfrak{q}})=0$.
Remembering $\Gamma>0,\; \mathfrak{\bar{q}}>0$ and $K>0$, by Descartes sign rule the equation admits a unique positive stationary solution. The solution has the property that, when $K/\Gamma\to 0$, $\mathfrak{q}^*\to\infty$. By linearizing the Eq.~\ref{eq:second-order-ode-nr} around its fixed point $(\mathfrak{q}=\mathfrak{q}^*, \dot{\mathfrak{q}}=0)$, we obtain the small oscillation frequency 
\begin{equation}\label{eq:freq}
\Omega = K^{1/2}\sqrt{1+\frac{2\Gamma}{K\mathfrak{q}^{*3}}}=K^{1/2}\sqrt{3-2\frac{\bar{\mathfrak{q}}}{\mathfrak{q}^*}}\,.
\end{equation}
One remarkable property of the oscillation frequency is the fact that, when $\mathfrak{q}^*\to\infty$, the frequency becomes $\Omega = \sqrt{3K}$.
Notice that this treatment is also valid in the full relativistic case, but is less physically significant, since we are considering small oscillations, and typical optomechanical experiments are in the nonrelativistic regime \cite{kippenbergAnalysisRadiationPressureInduced2005, marquardtDynamicalMultistabilityInduced2006}.
The position of the fixed point and the corresponding frequencies are represented in Fig.~\ref{fig:fixed_point}.
Panel~(a) displays the effective potential $\mathcal{V}(q)$ as a function of the coordinate $q$, together with its two main contributions: the radiation pressure term $\mathcal{V}_{\Gamma}$ and the elastic term $\mathcal{V}_{K}$. 
The combined potential $\mathcal{V}(q)$ results from the competition between these two contributions. 
Panel~(b) characterizes the equilibrium configuration by plotting the fixed-point coordinate $\mathfrak{q}^{*}$ (top) and the frequency of small oscillations $\Omega$ around this equilibrium (bottom), both as a function of the ratio $K/\Gamma$. 
The results show that increasing $K/\Gamma$ shifts the equilibrium position toward smaller displacements and reduces the oscillation frequency toward its background value.

The shift in frequency may be an experimentally accessible prediction.
In our nonrelativistic description, the shift depends only on the intracavity energy to stiffness ratio $\Gamma/K$. In experiments this energy is tuned by varying the detuning between the drive laser and the cavity resonance; accordingly, the measured shift in the oscillation frequency effect is usually presented as a function of that 
\cite{aspelmeyerCavityOptomechanics2014, kippenbergAnalysisRadiationPressureInduced2005}.
We remark that the known optical spring effect is related to the fact that the cavity becomes on and off resonant with respect to an external laser light \cite{corbittAllOpticalTrapGramScale2007, corbittMeasurementRadiationpressureinducedOptomechanical2006, sheardObservationCharacterizationOptical2004}. The spring constant of the optical spring \cb{described in previous works} can be extremely high \cb{with enhancement factors of order $10^3$ with respect to the bare mechanical frequency}. \cb{This } is related to the resonance condition, spanning in space only a few nanometers or less \cite{sheardObservationCharacterizationOptical2004, meystre1985theory}. The situation of our analysis \cb{complementary to the one obtained for the optical spring effect, as in our} cavity \cb{with } a single mode \cb{the resonance condition is always satisfied, and therefore the enhancement of the mechanical stiffness is less strong, as we show in the following}. 
Using data from the experiment reported in Ref.~\cite{corbittAllOpticalTrapGramScale2007}, consisting in a mass of $m=10^{-3}\,\mathrm{kg}$, an initial position of $q_0=0.9\,\mathrm{m}$, the bare mechanical resonance frequency $\Omega = 2\pi\times 172\,\mathrm{Hz}$, \cb{optical } resonance linewidth of $\gamma=2\pi\times 11\,\mathrm{kHz}$, and input power $P_{\mathrm{in}}=3\,\mathrm{W}$, and effective cavity gain $5\times10^3$, meaning that the circulating power is $P_{\mathrm{circ}}=gP_{\mathrm{in}}$, the stored energy is $E_0\approx 0.217\,\mathrm{J}$, which gives $\Gamma=E_0/(mc^2)\approx 2.4\times10^{-15}$ and $K=(\Omega q_0/c)^2\approx1.0\times10^{-11}$. Substituting in our expression for the resulting frequency (\ref{eq:freq}), we obtain a frequency shift of about $40\,\mathrm{mHz}$ for these cavity parameters. By increasing the power to $P_\mathrm{in}=30 \,$W, the frequency shift is modified to about $0.4\,\mathrm{Hz}$.

\subsection{Free mirror relativistic dynamics}\label{sec:libero}
\begin{figure}[t]
    \centering
    \begin{subfigure}[t]{0.49\textwidth}
    \includegraphics[width=\linewidth]{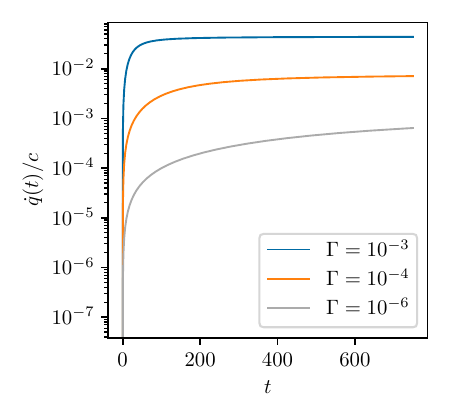}
    \caption{}
    \end{subfigure}
    \label{fig:wavelengths}
    \begin{subfigure}[t]{0.49\textwidth}
        \includegraphics[width=\linewidth]{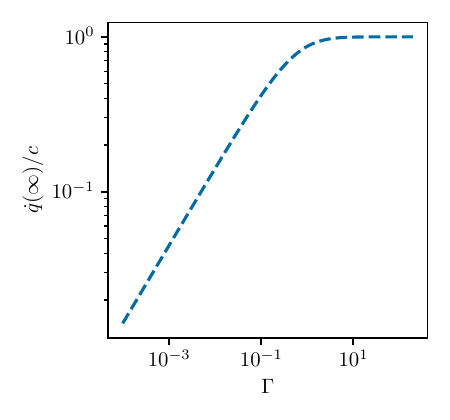}
        \caption{}
    \end{subfigure}%
    \caption{Solution of the equation of motion of the free mirror. Panel (a) shows the velocity over time for a range of radiation pressure parameters $\Gamma$. Panel (b) shows the limiting value of the final velocity as a function of the radiation pressure parameters. \cb{The initial velocity is set to zero. } $t$ is in units of $q_0/c$\cb{, $q_0$ being the initial position.}}
    \label{longdyne}
\end{figure}

In the special case of the absence of an external potential,  i.e. $K=0$,  the free motion of the mirror can be computed through the integral resulting from the substitution in Eq.~(\ref{eq:integralone}) of $\mathcal{E} - \mathcal{V}(s) = \Gamma\left(1-1/s\right)+1$. The integral can be solved, the result is the relation
\cb{
\begin{align}
t = \frac{(1+\Gamma)\sqrt{(\mathfrak{q}-1)(\mathfrak{q}(2+\Gamma)-\Gamma)}}{\Gamma^{1/2} (2+\Gamma)} + \frac{2}{\Gamma^{1/2}(2+\Gamma)^{3/2}}\operatorname{arcsinh}\left(\sqrt{\frac{(\mathfrak{q}-1)(2+\Gamma)}{2}} \right)
\end{align}
}
It is possible to consider the terminal velocity by differentiating on both sides, and taking the limit of $\mathfrak{q} \to \infty$, corresponding to the position at infinite time. The result is 
\cb{
\begin{equation}\label{eq:qinf}
    \dot{\mathfrak{q}}(\infty) = \frac{\sqrt{\Gamma (2 + \Gamma)}}{1 + \Gamma} \,,
\end{equation}
where we notice that the velocity cannot exceed the speed of light, since $\Gamma(2+\Gamma) < (1+\Gamma)^2$ for every $\Gamma$. }
The solutions for various $\Gamma$ and the terminal velocities are represented in Fig.~\ref{longdyne}. The figure illustrates the dynamics of the free mirror over time under the influence of radiation pressure.
Panel~(a) shows the time evolution of the mirror’s velocity $\dot{q}(t)$ for different values of the radiation pressure parameter $\Gamma$.
For small $\Gamma$, the velocity increases slowly and remains well below $c$ over the simulated timescale, while larger values of $\Gamma$ lead to a more rapid acceleration and earlier approach to the highly relativistic regime. 
Panel~(b) summarizes the asymptotic behavior by plotting the final velocity $\dot{q}(\infty)$ as a function of $\Gamma$. 
The plot shows a crossover from a power-law dependence at small $\Gamma$ to a saturation regime at large $\Gamma$. \cb{We remark that, } since we work within the approximation of \cb{small velocity } (see also the derivation of Eq.~\ref{eq:simplicio}), we consider reliable only the results obtained for low final velocities.
\cb{
Relativistic corrections become increasingly relevant as $\Gamma$ grows. To assess their impact on the terminal velocity, we also solved numerically the corresponding nonrelativistic equation, namely Eq.~\eqref{eq:second-order-ode} with the first factor on the right-hand side approximated by unity. We calculate the relative error $\varepsilon_r = \lvert \dot{\mathfrak{q}}_{\mathrm{nr}}(\infty) - \dot{\mathfrak{q}}(\infty) \rvert / \dot{\mathfrak{q}}(\infty)$, where $\dot{\mathfrak{q}}_{\mathrm{nr}}(\infty)$ is the one obtained in the nonrelativistic case. For conservative values of $\Gamma$, it is at the level of a few percent: for instance, for $\Gamma = 3.16\times10^{-2}$ we have $\varepsilon_r = 2.18\times10^{-2}$; and for $\Gamma = 1.00\times10^{-1}$ we have $\varepsilon_r = 7.07\times10^{-2}$. The nonrelativistic calculation overestimate the terminal velocity. 
To estimate the order of magnitudes of the physical parameters which make the relativistic corrections relevant, we can use the parameters of the interstellar--flight scenario discussed in Ref.~\cite{lorenzi-prapplied}. For the reference launch protocol, the lightsail mass is $m=10~\mathrm{g}$, the laser power is $P=50~\mathrm{GW}$, and the thrust duration is $t_{\mathrm F}=500~\mathrm{s}$, so that the total laser energy delivered during the launch is approximately $P t_{\mathrm F}\approx 2.5\times10^{13}\,\mathrm{J}$. We can relate this quantity to the upper bound to the parameter $\Gamma$. In the case in which the energy calculated above is fully stored in a single mode, we obtain a maximum value $\Gamma_{\mathrm{max}} \approx 2.8\times10^{-2}$. Thus, even in the case of an interstellar flight, the dimensionless driving parameter remains in a regime where the relativistic corrections are of a few--percent level. In this simplified case, the final velocity as per Eq.~\eqref{eq:qinf} is $\dot{q}(\infty)\approx 0.23c$, and, by considering an initial distance of $q_0=3.5\times 10^7$ m,  it is reached in a time scale of hundreds of seconds. We remark however that the model developed in this work bears substantial differences to the one reported in Ref.~\cite{lorenzi-prapplied}, and further investigations are needed to assess the relation between them.
}

\section{Discussion}
The results obtained here demonstrate that the dynamics of the mirror can, to a good extent, be analyzed independently of the full dynamics of the field, provided that suitable approximations are made.  
The key assumption underlying this separation is that it is possible to perform an adiabatic elimination of the electromagnetic field. This idea is inspired by the early work of Baranov and Shirokov, who showed that a moving-mirror boundary condition gives rise to dynamic modes in the cavity, where Doppler-shift–induced modulation of the field spectrum \cite{baranov1967electromagnetic} changes the mode structure. In this picture, the mirror motion directly reshapes the cavity resonance structure, leading to deviations from the monochromatic modes of a static cavity, but the calculations are limited to the case of uniform motion.
By treating the mirror as a dynamical degree of freedom, other approaches explicitly showed how radiation pressure directly enters the dynamical equations of the cavity modes inducing an interaction between them \cite{law1995interaction, razavyQuantumRadiationOnedimensional1985}. The two viewpoints are consistent: the former emphasizes the kinematic modification of mode structure, while the latter establishes a canonical formalism suited to quantization. However, they are not directly mappable to one another. Our variational single-mode approach, leading to the reduced equation~(\ref{eq:second-order-ode}), inherits aspects of both pictures. It can predict dynamical features such as the shifted stationary point, the frequency of small oscillations, and the motion of a free mirror in the weakly relativistic regime.

Because we assume a single cavity mode in isolation, the treatment cannot describe phenomena where multimode excitation and external pumping are essential, such as optical bistability, parametric instabilities due to dynamical backaction, and radiation-pressure–induced dissipation \cite{braginskiiInvestigation1970, braginskyParametricOscillatoryInstability2001, kippenbergCavityOptomechanicsBackAction2008a, aspelmeyerCavityOptomechanics2014, meystre1985theory}. These effects require input fields that drive the cavity off resonance, as in interferometers and optomechanical setups \cite{aspelmeyerCavityOptomechanics2014,kippenbergCavityOptomechanicsBackAction2008a, corbittAllOpticalTrapGramScale2007}. By contrast, in our framework the selected mode is always somehow resonant with the moving boundary.

Another central assumption is energy conservation over the relevant timescales. This is justified in high--quality-factor cavities, where losses due to absorption, scattering, or non-ideal reflectivity only become significant after many round trips.  
The quality factor is defined as $Q = \omega U/P_{\mathrm{loss}}$, with $\omega$ the mode frequency, $U$ the stored energy, and $P_{\mathrm{loss}}$ the dissipated power \cite{salehFundamentalsPhotonics22019}. For a Fabry--Pérot cavity of length $L$ and mirror reflectivity $R$, one finds $Q \sim 2 \pi \ell L \sqrt{R}/(\lambda(1-R))$. Thus, at fixed $R$, the quality factor grows linearly with $L$, although in practice it eventually decreases for very long cavities due to residual absorption and scattering. Our conservative single-mode model is therefore most accurate for large-$Q$ cavities with extremely high mirror reflectivities.
In realistic systems, however, an external pump is needed to replenish cavity losses. Pumping introduces input--output coupling and the possibility of populating several modes, thereby modifying the mirror dynamics in nontrivial ways. 

%
%
\section{Conclusions}

In this work, we have developed a variational framework to study the dynamics of an optical cavity with a moving mirror, treating both relativistic and nonrelativistic regimes. Rather than solving the full set of coupled equations for the electromagnetic field and mirror motion, we employed a single-mode variational ansatz that captures the essential physics of radiation–mirror interaction while remaining analytically tractable. This method provides a clear and intuitive picture of how the cavity field and mirror dynamics are intrinsically coupled.
By applying the WKB approximation in the short-wavelength limit, we derived effective equations of motion for the mirror and identified the emergence of stationary points as well as oscillatory behavior. In the limiting case of a free mirror, we obtained exact analytic solutions and showed that radiation pressure drives the mirror toward a finite terminal velocity determined by the initial energy stored in the cavity.

Overall, our results provide a treatable approximation to the dynamics of classical optomechanical cavity. The framework presented in this work offers a theoretical approach for future studies of radiation-pressure propulsion concepts, including cavity-assisted thrusters and laser-driven lightsails, as well as precision interferometric experiments where mirror motion plays a decisive role across different dynamical regimes.
Extending the present treatment to include multimode interactions and external pumping is a natural direction for future work. Such extensions are crucial to connect this simplified picture with experimental realizations in cavity optomechanics, laser interferometry, and related platforms.
\cb{In addition, a quantitative analysis of dissipation and environmental effects, in particular non-ideal mirror reflectivity and mirror damping, would clarify the regimes where the current approximations remain valid. Incorporating these mechanisms would also enable a more realistic comparison with existing experiments.
Finally, a detailed study of quantum fluctuations of both the mirror and the field would deepen the understanding of the system’s behavior and guide future experimental investigations.
}
\section*{Data availability statement}
No new data were created or analysed in the study.
\cb{
\section*{Acknowledgements}
M.G.P. acknowledges Progetto Space-it Up!, Bando (Prot. CI-2022-DSR-042) per le “Attività spaziali” (tematica 15), MUR n. 341 del 15/03/2022.
}
\section*{Appendix A: Bounded oscillations in the relativistic regime}
\begin{figure}[h]
    \centering

    \begin{subfigure}[t]{0.49\textwidth}
        \includegraphics[width=\linewidth]{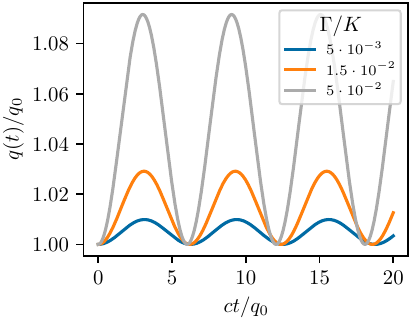}
        \caption{}
        \label{fig:sympl_NR_q}
    \end{subfigure}
    \hfill
    \begin{subfigure}[t]{0.49\textwidth}
        \includegraphics[width=\linewidth]{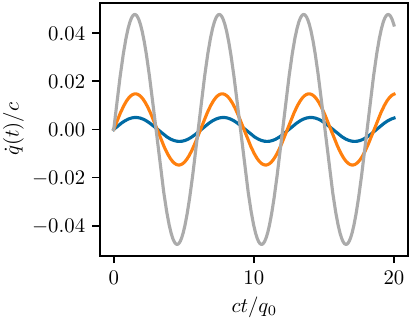}
        \caption{}
        \label{fig:sympl_NR_qdot}
    \end{subfigure}

    \vspace{0.8em}

    \begin{subfigure}[t]{0.49\textwidth}
        \includegraphics[width=\linewidth]{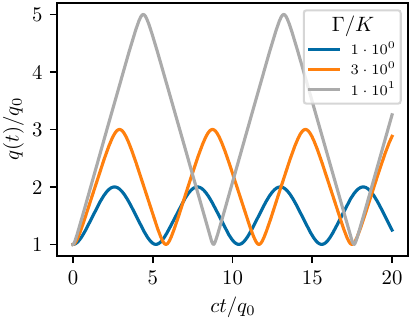}
        \caption{}
        \label{fig:sympl_rel_q}
    \end{subfigure}
    \hfill
    \begin{subfigure}[t]{0.49\textwidth}
        \includegraphics[width=\linewidth]{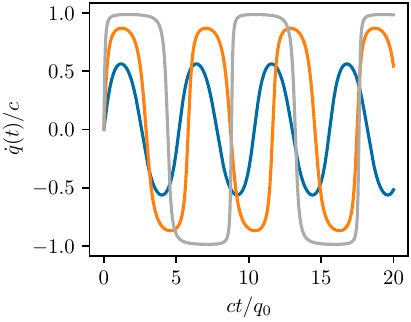}
        \caption{}
        \label{fig:sympl_rel_qdot}
    \end{subfigure}

    \caption{\cb{Time evolution of the mirror’s position $q$ and velocity $\dot{q}$ obtained from the equation of motion (Eq.~\ref{eq:second-order-ode}) for different ratios of radiation pressure to elastic coefficients, $\Gamma/K$, with $K=1$. The upper panels~(a)–(b) correspond to the nonrelativistic regime, whereas the lower panels~(c)–(d) show the relativistic case. The initial velicity is set to zero, and the initial position is $q_0$.}}
    \label{fig:sympl}
\end{figure}

The case of a pure relativistic harmonic oscillator, i.e. the case for which Eq.~(\ref{eq:second-order-ode}) has been studied in previous works \cite{babusciRelativisticHarmonicOscillator2013, zarmiRelativisticHarmonicOscillator2023}, and can be analyzed in terms of Jacobi elliptic functions \cite{moreauRelativisticAnharmonicOscillator1994}.
The case of relativistic motion within a harmonic potential is less relevant from a physical perspective, but it presents mathematically interesting properties. For the sake of completeness, we report here the results obtained in the case of a relativistic motion in a harmonic potential under the pressure of radiation.
Figure~\ref{fig:sympl} presents the solutions of the mirror’s equation of motion for different ratios of radiation pressure to elastic restoring force ($\Gamma/K$), with the stiffness fixed at $K = 1$. \cb{The initial conditions are $q(0)=q_0$, and $\dot{q}(0)=0$. }
Panel~(a) shows the mirror position $q(t)$ as a function of time, while panel~(b) shows the corresponding velocity $\dot{q}(t)$.
For small values of $\Gamma/K$, the oscillations remain nearly sinusoidal, indicating weak distortion by the radiation pressure term.
As $\Gamma/K$ increases, however, the dynamics becomes strongly nonlinear: the position and velocity exhibit asymmetric, non-sinusoidal oscillations with sharper features and anharmonic profiles. 
This illustrates the progressive dominance of radiation pressure over the elastic response in shaping the motion of the mirror. \cb{
The presence of the radiation pressure term induce a static bias and a small correction to the natural frequency of the relativistic harmonic oscillator, without altering the fundamental relativistic character of the motion. The optical force shifts the equilibrium position of the oscillator, establishing a new point around which the relativistic dynamics take place.  In contrast, the relativistic kinematics still governs the amplitude–dependent deviation from harmonicity of the motion. Thus, the radiation pressure acts as a perturbative correction to the equilibrium and to the linear elastic constant, while the relativistic term remains responsible for the nonlinear frequency shift and the deformation of the periodic orbit. 
As shown in Babusci  \cite{babusciRelativisticHarmonicOscillator2013}, if one keeps only the fist relativistic correction to the classical motion of a relativistic harmonic oscillator, one gets a Duffing-type dynamical equation \cite[Eq. (3.2)]{babusciRelativisticHarmonicOscillator2013}. In this context, the radiation pressure could be studied perturbatively using well known methods.
}
It should be noted that the plotted velocity corresponds to the relativistic velocity normalized to the speed of light $c$; in the highly relativistic regime, the present model is expected to lose validity and a more complete relativistic treatment would be required.

\cb{
\section*{Appendix B: Mirror effective dynamics in presence of quantum and thermal fluctuations}
Our nonrelativistic equation of motion for the mirror coordinate, Eq.~\eqref{eq:second-order-ode-nr}, obtaines with the adiabatic elimination of the interaction with a single-mode, can be derived from a corresponding classical Lagrangian of the shape
\begin{equation}
  L=\frac{m}{2}\dot q^{\,2}-\mathcal{V}(q).
\end{equation}
where $\mathcal{V}$ is the effective potential given by Eq.~\eqref{eq:potential}.
The Lagrangian can be utilized to derive within a quantum effective potential for a system of bosons \cite{jona-lasinioRelativisticFieldTheories1964, colemanRadiativeCorrectionsOrigin1973, goldstoneBrokenSymmetries1962, furutaniQuantumEffectiveAction2022d, burgess2020introduction}. Defining the effective frequency $\omega(q) = \sqrt{\mathcal{V}''(q)/m}$, the one–loop effective potential, composed by the contribution of zero-point fluctuations and thermal fluctuations is
\begin{equation}
  \mathcal{V}_{\mathrm{eff}}(q)=\mathcal{V}(q)+\tilde{\mathcal{V}}_0(q)+\tilde{\mathcal{V}}_T(q),
\end{equation}
with the contribution of zero-point energy
\begin{equation}
  \tilde{\mathcal{V}}_0(q)=\frac{\hbar}{2}\,\omega(q)
  =\frac{\hbar}{2}\sqrt{\frac{\mathcal{V}''(q)}{m}},
\end{equation}
and thermal energy
\begin{equation}
      \tilde{\mathcal{V}}_T(q)=k_B T\,
  \ln \left[1-\exp\!\Big(-\frac{\hbar\,\omega(q)}{k_B T}\Big)\right].
\end{equation}
with $T$ the absolute temperature of the system and $k_B$ the Boltzmann constant.
Utilizing the normalized mirror coordinate $\mathfrak{q}$ and explicitly writing the potential, we have
\begin{equation}
  \mathcal{V}''(\mathfrak{q})=K+\frac{2\Gamma}{\mathfrak{q}^3} \,,
\end{equation}
%
This formalism allows to perturbatively compute the deviation to the oscillation frequency in presence of nonzero temperature  and quantum fluctuations. We consider a temperature of $T=300$K. Using the same parameters as in Section~\ref{sec:molla}, and assuming that the stationary position $q^*$ is unchanged, the calculated shift in the frequency due to quantum and thermal fluctuations are many order of magnitudes lower than the one predicted in Section~\ref{sec:molla}, making this effect negligible. However, we point out that this treatment is unable to describe the interaction with the external environment through dissipative processes.
}

\FloatBarrier
\printbibliography

@article{bae2021photonic,
  title={Photonic laser thruster: 100 times scaling-up and propulsion demonstration},
  author={Bae, Young K},
  journal={Journal of Propulsion and Power},
  volume={37},
  number={3},
  pages={400--407},
  year={2021},
  publisher={American Institute of Aeronautics and Astronautics}
}

@article{bae2022photonic,
  title={Photonic Laser Thruster: Optomechanical and Quantum Electronical Analyses},
  author={Bae, Young K},
  journal={Journal of Propulsion and Power},
  volume={38},
  number={3},
  pages={437--449},
  year={2022},
  publisher={American Institute of Aeronautics and Astronautics}
}

@article{baranov1967electromagnetic,
  title={Electromagnetic field in an Optical resonator with a movable mirror},
  author={Baranov, RI and Shirokov, Yu M},
  journal={Soviet Physics JETP},
  volume={53},
  pages={2123},
  year={1967}
}

@article{law1995interaction,
  title={Interaction between a moving mirror and radiation pressure: A Hamiltonian formulation},
  author={Law, CK},
  journal={Physical Review A},
  volume={51},
  number={3},
  pages={2537},
  year={1995},
  publisher={APS}
}

@article{moore1970quantum,
  title={Quantum theory of the electromagnetic field in a variable-length one-dimensional cavity},
  author={Moore, Gerald T},
  journal={Journal of Mathematical Physics},
  volume={11},
  number={9},
  pages={2679--2691},
  year={1970},
  publisher={American Institute of Physics}
}

@article{kulkarni2018relativistic,
  title={Relativistic spacecraft propelled by directed energy},
  author={Kulkarni, Neeraj and Lubin, Philip and Zhang, Qicheng},
  journal={The Astronomical Journal},
  volume={155},
  number={4},
  pages={155},
  year={2018},
  publisher={IOP Publishing}
}

@article{Law2012,
  title={Optomechanical coupling between a moving dielectric sphere and
radiation fields: a Lagrangian-Hamiltonian formalism},
  author={Cheung, H.K. and Law, C.K.},
  journal={Physical Review A},
  volume={86},
  number={033807},
  pages={},
  year={2012},
  publisher={}
}

@article{davoyan_photonic_2021,
	title = {Photonic materials for interstellar solar sailing},
	volume = {8},
	copyright = {\&\#169; 2021 Optical Society of America},
	issn = {2334-2536},
	url = {https://opg.Optica.org/Optica/abstract.cfm?uri=Optica-8-5-722},
	doi = {10.1364/Optica.417007},
	number = {5},
	urldate = {2025-02-27},
	journal = {Optica},
	author = {Davoyan, Artur R. and Munday, Jeremy N. and Tabiryan, Nelson and Swartzlander, Grover A. and Johnson, Les},
	month = may,
	year = {2021},
	note = {Publisher: Optica Publishing Group},
	pages = {722--734}
}

@article{nichols1903pressure,
  title={The pressure due to radiation.(second paper.)},
  author={Nichols, Ernest Fox and Hull, Gordon Ferrie},
  journal={Physical Review (Series I)},
  volume={17},
  number={1},
  pages={26},
  year={1903},
  publisher={APS}
}

@article{nichols1901preliminary,
  title={A preliminary communication on the pressure of heat and light radiation},
  author={Nichols, Ernest Fox and Hull, Gordon Ferrie},
  journal={Physical Review (Series I)},
  volume={13},
  number={5},
  pages={307},
  year={1901},
  publisher={APS}
}

@article{lebedev,
author = {Lebedew, Peter},
title = {Untersuchungen über die Druckkräfte des Lichtes},
journal = {Annalen der Physik},
volume = {311},
number = {11},
pages = {433-458},
doi = {https://doi.org/10.1002/andp.19013111102},
url = {https://onlinelibrary.wiley.com/doi/abs/10.1002/andp.19013111102},
eprint = {https://onlinelibrary.wiley.com/doi/pdf/10.1002/andp.19013111102},
year = {1901}
}

@article{ashkin1972pressure,
  title={The pressure of laser light},
  author={Ashkin, Arthur},
  journal={Scientific American},
  volume={226},
  number={2},
  pages={62--71},
  year={1972},
  publisher={JSTOR}
}

@article{calucci1992casimir,
  title={Casimir effect for moving bodies},
  author={Calucci, Giorgio},
  journal={Journal of Physics A: Mathematical and General},
  volume={25},
  number={13},
  pages={3873},
  year={1992},
  publisher={IOP Publishing}
}

@article{meystre1985theory,
  title={Theory of radiation-pressure-driven interferometers},
  author={Meystre, Pierre and Wright, Ewan M and McCullen, JD and Vignes, E},
  journal={Journal of the Optical Society of America B},
  volume={2},
  number={11},
  pages={1830--1840},
  year={1985},
  publisher={Optical Society of America}
}

@article{babusciRelativisticHarmonicOscillator2013,
  title = {Relativistic Harmonic Oscillator, the Associated Equations of Motion, and Algebraic Integration Methods},
  author = {Babusci, D. and Dattoli, G. and Quattromini, M. and Sabia, E.},
  year = {2013},
  month = mar,
  journal = {Physical Review E},
  volume = {87},
  number = {3},
  pages = {033202},
  publisher = {American Physical Society},
  doi = {10.1103/PhysRevE.87.033202},
  urldate = {2025-08-22}
}

@article{moreauRelativisticAnharmonicOscillator1994,
  title = {Relativistic (an)Harmonic Oscillator},
  author = {Moreau, William and Easther, Richard and Neutze, Richard},
  year = {1994},
  month = jun,
  journal = {American Journal of Physics},
  volume = {62},
  number = {6},
  pages = {531--535},
  issn = {0002-9505},
  doi = {10.1119/1.17513},
  urldate = {2025-08-22}
}

@article{zarmiRelativisticHarmonicOscillator2023,
  title = {On the {{Relativistic Harmonic Oscillator}}},
  author = {Zarmi, Yair},
  year = {2023},
  month = jan,
  journal = {Applied Mathematics},
  volume = {14},
  number = {1},
  pages = {1--20},
  publisher = {Scientific Research Publishing},
  doi = {10.4236/am.2023.141001},
  urldate = {2025-08-22},
  copyright = {http://creativecommons.org/licenses/by/4.0/},
  langid = {english}
}

@book{geometric-numerical,
title={Geometric Numerical
Integration,
Structure-Preserving
Algorithms for Ordinary
Differential Equations},
author={E. Hairer,
C. Lubich,
G. Wanner},
year={2006},
publisher={Springer},
}

@article{lorenzi-prapplied,
  title={Optical cavity in the relativistic regime for laser propulsion},
  author={Lorenzi, Francesco and Salasnich, Luca and Pelizzo, Maria Guglielmina},
  journal={Physical Review Applied},
  volume={24},
  number={3},
  pages={034033},
  year={2025},
  publisher={APS}, doi={https://doi.org/10.1103/pkn5-9mb4}
}

@article{aspelmeyerCavityOptomechanics2014,
  title = {Cavity Optomechanics},
  author = {Aspelmeyer, Markus and Kippenberg, Tobias J. and Marquardt, Florian},
  year = {2014},
  month = dec,
  journal = {Rev. Mod. Phys.},
  volume = {86},
  number = {4},
  pages = {1391--1452},
  publisher = {American Physical Society},
  doi = {10.1103/RevModPhys.86.1391},
  urldate = {2025-09-09}
}

@article{braginski1967ponderomotive,
  title={Ponderomotive effects of electromagnetic radiation},
  author={Braginski, VB and Manukin, AB},
  journal={Sov. Phys. JETP},
  volume={25},
  number={4},
  pages={653--655},
  year={1967}
}

@article{braginskyLowQuantumNoise2002,
  title = {Low Quantum Noise Tranquilizer for {{Fabry}}--{{Perot}} Interferometer},
  author = {Braginsky, V. B. and Vyatchanin, S. P.},
  year = {2002},
  month = feb,
  journal = {Physics Letters A},
  volume = {293},
  number = {5},
  pages = {228--234},
  issn = {0375-9601},
  doi = {10.1016/S0375-9601(02)00020-8},
  urldate = {2025-09-09}
}

@article{braginskyParametricOscillatoryInstability2001,
  title = {Parametric Oscillatory Instability in {{Fabry}}--{{Perot}} Interferometer},
  author = {Braginsky, V. B. and Strigin, S. E. and Vyatchanin, S. P.},
  year = {2001},
  month = sep,
  journal = {Physics Letters A},
  volume = {287},
  number = {5},
  pages = {331--338},
  issn = {0375-9601},
  doi = {10.1016/S0375-9601(01)00510-2},
  urldate = {2025-09-09}
}

@article{jirauschekWavelengthShiftingIntracavity2015,
  title = {Wavelength Shifting of Intra-Cavity Photons: {{Adiabatic}} Wavelength Tuning in Rapidly Wavelength-Swept Lasers},
  shorttitle = {Wavelength Shifting of Intra-Cavity Photons},
  author = {Jirauschek, Christian and Huber, Robert},
  year = {2015},
  month = jul,
  journal = {Biomedical Optics Express},
  volume = {6},
  number = {7},
  pages = {2448--2465},
  publisher = {Optica Publishing Group},
  issn = {2156-7085},
  doi = {10.1364/BOE.6.002448},
  urldate = {2024-11-13},
  copyright = {{\copyright} 2015 Optical Society of America},
  langid = {english}
}

@article{kippenbergAnalysisRadiationPressureInduced2005,
  title = {Analysis of {{Radiation-Pressure Induced Mechanical Oscillation}} of an {{Optical Microcavity}}},
  author = {Kippenberg, T. J. and Rokhsari, H. and Carmon, T. and Scherer, A. and Vahala, K. J.},
  year = {2005},
  month = jul,
  journal = {Physical Review Letters},
  volume = {95},
  number = {3},
  pages = {033901},
  publisher = {American Physical Society},
  doi = {10.1103/PhysRevLetters95.033901},
  urldate = {2025-09-09}
}

@article{kranendonkModelessOperationWavelengthagile2005,
  title = {Modeless Operation of a Wavelength-Agile Laser by High-Speed Cavity Length Changes},
  author = {Kranendonk, Laura A. and Bartula, Renata J. and Sanders, Scott T.},
  year = {2005},
  month = mar,
  journal = {Optics Express},
  volume = {13},
  number = {5},
  pages = {1498--1507},
  publisher = {Optica Publishing Group},
  issn = {1094-4087},
  doi = {10.1364/OPEX.13.001498},
  urldate = {2024-11-13},
  copyright = {{\copyright} 2005 Optical Society of America},
  langid = {english}
}

@article{marquardtDynamicalMultistabilityInduced2006,
  title = {Dynamical {{Multistability Induced}} by {{Radiation Pressure}} in {{High-Finesse Micromechanical Optical Cavities}}},
  author = {Marquardt, Florian and Harris, J. G. E. and Girvin, S. M.},
  year = {2006},
  month = mar,
  journal = {Physical Review Letters},
  volume = {96},
  number = {10},
  pages = {103901},
  publisher = {American Physical Society},
  doi = {10.1103/PhysRevLetters96.103901},
  urldate = {2025-09-09}
}

@article{marquardtQuantumTheoryCavityAssisted2007,
  title = {Quantum {{Theory}} of {{Cavity-Assisted Sideband Cooling}} of {{Mechanical Motion}}},
  author = {Marquardt, Florian and Chen, Joe P. and Clerk, A. A. and Girvin, S. M.},
  year = {2007},
  month = aug,
  journal = {Physical Review Letters},
  volume = {99},
  number = {9},
  pages = {093902},
  publisher = {American Physical Society},
  doi = {10.1103/PhysRevLetters99.093902},
  urldate = {2025-09-09}
}

@article{meystreShortWalkQuantum2013,
  title = {A Short Walk through Quantum Optomechanics},
  author = {Meystre, Pierre},
  year = {2013},
  journal = {Annalen der Physik},
  volume = {525},
  number = {3},
  pages = {215--233},
  issn = {1521-3889},
  doi = {10.1002/andp.201200226},
  urldate = {2025-09-09},
  langid = {english}
}

@article{notomiWavelengthConversionDynamic2006,
  title = {Wavelength Conversion via Dynamic Refractive Index Tuning of a Cavity},
  author = {Notomi, Masaya and Mitsugi, Satoshi},
  year = {2006},
  month = may,
  journal = {Physical Review A},
  volume = {73},
  number = {5},
  pages = {051803},
  publisher = {American Physical Society},
  doi = {10.1103/PhysRevA.73.051803},
  urldate = {2024-11-13}
}

@article{prebleSinglePhotonAdiabatic2012,
  title = {Single Photon Adiabatic Wavelength Conversion},
  author = {Preble, Stefan and Cao, Liang and Elshaari, Ali and Aboketaf, Abdelsalam and Adams, Donald},
  year = {2012},
  month = oct,
  journal = {Applied Physics Letters},
  volume = {101},
  number = {17},
  pages = {171110},
  issn = {0003-6951},
  doi = {10.1063/1.4764068},
  urldate = {2024-11-13}
}

@article{wilson-raeTheoryGroundState2007,
  title = {Theory of {{Ground State Cooling}} of a {{Mechanical Oscillator Using Dynamical Backaction}}},
  author = {{Wilson-Rae}, I. and Nooshi, N. and Zwerger, W. and Kippenberg, T. J.},
  year = {2007},
  month = aug,
  journal = {Physical Review Letters},
  volume = {99},
  number = {9},
  pages = {093901},
  publisher = {American Physical Society},
  doi = {10.1103/PhysRevLetters99.093901},
  urldate = {2025-09-09}
}

@article{kippenbergCavityOptomechanicsBackAction2008a,
  title = {Cavity {{Optomechanics}}: {{Back-Action}} at the {{Mesoscale}}},
  shorttitle = {Cavity {{Optomechanics}}},
  author = {Kippenberg, T. J. and Vahala, K. J.},
  year = {2008},
  month = aug,
  journal = {Science},
  volume = {321},
  number = {5893},
  pages = {1172--1176},
  publisher = {American Association for the Advancement of Science},
  doi = {10.1126/science.1156032},
  urldate = {2025-09-19}
}

@article{braginskiiInvestigation1970,
  title = {Investigation of {{Dissipative Ponderomotive Effects}} of {{Electromagnetic Radiation}}},
  author = {Braginski{\v i}, V. B. and Manukin, A. B. and Tikhonov, M. {\relax Yu}.},
  year = {1970},
  month = jan,
  journal = {Soviet Journal of Experimental and Theoretical Physics},
  volume = {31},
  pages = {829},
  issn = {1063-7761},
  urldate = {2025-09-19}
}

@article{lawEffectiveHamiltonianRadiation1994a,
  title = {Effective {{Hamiltonian}} for the Radiation in a Cavity with a Moving Mirror and a Time-Varying Dielectric Medium},
  author = {Law, C. K.},
  year = {1994},
  journal = {Physical Review A},
  volume = {49},
  number = {1},
  pages = {433--437},
  publisher = {American Physical Society},
  doi = {10.1103/PhysRevA.49.433},
  urldate = {2025-09-22}
}

@article{razavyQuantumRadiationOnedimensional1985,
  title = {Quantum Radiation in a One-Dimensional Cavity with Moving Boundaries},
  author = {Razavy, M. and Terning, J.},
  year = {1985},
  journal = {Physical Review D},
  volume = {31},
  number = {2},
  pages = {307--313},
  publisher = {American Physical Society},
  doi = {10.1103/PhysRevD.31.307},
  urldate = {2025-09-22}
}

@book{salehFundamentalsPhotonics22019,
  title = {Fundamentals of {{Photonics}}, 2 {{Volume Set}}},
  author = {Saleh, Bahaa E. A. and Teich, Malvin Carl},
  year = {2019},
  publisher = {John Wiley \& Sons},
  googlebooks = {urMzEQAAQBAJ},
  isbn = {978-1-119-50687-4},
  langid = {english}
}

@article{corbittAllOpticalTrapGramScale2007,
  title = {An {{All-Optical Trap}} for a {{Gram-Scale Mirror}}},
  author = {Corbitt, Thomas and Chen, Yanbei and Innerhofer, Edith and {M{\"u}ller-Ebhardt}, Helge and Ottaway, David and Rehbein, Henning and Sigg, Daniel and Whitcomb, Stanley and Wipf, Christopher and Mavalvala, Nergis},
  year = {2007},
  month = apr,
  journal = {Physical Review Letters},
  volume = {98},
  number = {15},
  pages = {150802},
  publisher = {American Physical Society},
  doi = {10.1103/PhysRevLett.98.150802},
  urldate = {2025-09-23}
}

@article{corbittMeasurementRadiationpressureinducedOptomechanical2006,
  title = {Measurement of Radiation-Pressure-Induced Optomechanical Dynamics in a Suspended {{Fabry-Perot}} Cavity},
  author = {Corbitt, Thomas and Ottaway, David and Innerhofer, Edith and Pelc, Jason and Mavalvala, Nergis},
  year = {2006},
  month = aug,
  journal = {Physical Review A},
  volume = {74},
  number = {2},
  pages = {021802},
  publisher = {American Physical Society},
  doi = {10.1103/PhysRevA.74.021802},
  urldate = {2025-09-23}
}

@article{sheardObservationCharacterizationOptical2004,
  title = {Observation and Characterization of an Optical Spring},
  author = {Sheard, Benjamin S. and Gray, Malcolm B. and {Mow-Lowry}, Conor M. and McClelland, David E. and Whitcomb, Stanley E.},
  year = {2004},
  month = may,
  journal = {Physical Review A},
  volume = {69},
  number = {5},
  pages = {051801},
  publisher = {American Physical Society},
  doi = {10.1103/PhysRevA.69.051801},
  urldate = {2025-09-23}
}

@article{atwaterMaterialsChallengesStarshot2018f,
  title = {Materials Challenges for the {{Starshot}} Lightsail},
  author = {Atwater, Harry A. and Davoyan, Artur R. and Ilic, Ognjen and Jariwala, Deep and Sherrott, Michelle C. and Went, Cora M. and Whitney, William S. and Wong, Joeson},
  year = {2018},
  month = oct,
  journal = {Nature Materials},
  volume = {17},
  number = {10},
  pages = {861--867},
  publisher = {Nature Publishing Group},
  issn = {1476-4660},
  doi = {10.1038/s41563-018-0075-8},
  urldate = {2025-09-25},
  copyright = {2018 Springer Nature Limited},
  langid = {english}
}

@article{daniaHighpurityQuantumOptomechanics2025,
  title = {High-Purity Quantum Optomechanics at Room Temperature},
  author = {Dania, Lorenzo and Kremer, Oscar Schmitt and Piotrowski, Johannes and Candoli, Davide and Vijayan, Jayadev and {Romero-Isart}, Oriol and {Gonzalez-Ballestero}, Carlos and Novotny, Lukas and Frimmer, Martin},
  year = {2025},
  month = aug,
  journal = {Nature Physics},
  pages = {1--6},
  publisher = {Nature Publishing Group},
  issn = {1745-2481},
  doi = {10.1038/s41567-025-02976-9},
  urldate = {2025-09-25},
  copyright = {2025 The Author(s)},
  langid = {english}
}

@article{groblacherDemonstrationUltracoldMicrooptomechanical2009,
  title = {Demonstration of an Ultracold Micro-Optomechanical Oscillator in a Cryogenic Cavity},
  author = {Gr{\"o}blacher, Simon and Hertzberg, Jared B. and Vanner, Michael R. and Cole, Garrett D. and Gigan, Sylvain and Schwab, K. C. and Aspelmeyer, Markus},
  year = {2009},
  month = jul,
  journal = {Nature Physics},
  volume = {5},
  number = {7},
  pages = {485--488},
  publisher = {Nature Publishing Group},
  issn = {1745-2481},
  doi = {10.1038/nphys1301},
  urldate = {2025-09-25},
  copyright = {2009 Springer Nature Limited},
  langid = {english}
}

@article{huangRoomtemperatureQuantumOptomechanics2024,
  title = {Room-Temperature Quantum Optomechanics Using an Ultralow Noise Cavity},
  author = {Huang, Guanhao and Beccari, Alberto and Engelsen, Nils J. and Kippenberg, Tobias J.},
  year = {2024},
  month = feb,
  journal = {Nature},
  volume = {626},
  number = {7999},
  pages = {512--516},
  publisher = {Nature Publishing Group},
  issn = {1476-4687},
  doi = {10.1038/s41586-023-06997-3},
  urldate = {2025-09-25},
  copyright = {2024 The Author(s)},
  langid = {english}
}

@article{ilicSelfstabilizingPhotonicLevitation2019e,
  title = {Self-Stabilizing Photonic Levitation and Propulsion of Nanostructured Macroscopic Objects},
  author = {Ilic, Ognjen and Atwater, Harry A.},
  year = {2019},
  month = apr,
  journal = {Nature Photonics},
  volume = {13},
  number = {4},
  pages = {289--295},
  publisher = {Nature Publishing Group},
  issn = {1749-4893},
  doi = {10.1038/s41566-019-0373-y},
  urldate = {2025-09-25},
  copyright = {2019 The Author(s), under exclusive licence to Springer Nature Limited},
  langid = {english}
}

@article{riedingerNonclassicalCorrelationsSingle2016,
  title = {Non-Classical Correlations between Single Photons and Phonons from a Mechanical Oscillator},
  author = {Riedinger, Ralf and Hong, Sungkun and Norte, Richard A. and Slater, Joshua A. and Shang, Juying and Krause, Alexander G. and Anant, Vikas and Aspelmeyer, Markus and Gr{\"o}blacher, Simon},
  year = {2016},
  month = feb,
  journal = {Nature},
  volume = {530},
  number = {7590},
  pages = {313--316},
  publisher = {Nature Publishing Group},
  issn = {1476-4687},
  doi = {10.1038/nature16536},
  urldate = {2025-09-25},
  copyright = {2016 Springer Nature Limited},
  langid = {english}
}

@article{vannerCoolingbymeasurementMechanicalState2013,
  title = {Cooling-by-Measurement and Mechanical State Tomography via Pulsed Optomechanics},
  author = {Vanner, M. R. and Hofer, J. and Cole, G. D. and Aspelmeyer, M.},
  year = {2013},
  month = aug,
  journal = {Nature Communications},
  volume = {4},
  number = {1},
  pages = {2295},
  publisher = {Nature Publishing Group},
  issn = {2041-1723},
  doi = {10.1038/ncomms3295},
  urldate = {2025-09-25},
  copyright = {2013 Springer Nature Limited},
  langid = {english}
}

@article{verhagenQuantumcoherentCouplingMechanical2012,
  title = {Quantum-Coherent Coupling of a Mechanical Oscillator to an Optical Cavity Mode},
  author = {Verhagen, E. and Del{\'e}glise, S. and Weis, S. and Schliesser, A. and Kippenberg, T. J.},
  year = {2012},
  month = feb,
  journal = {Nature},
  volume = {482},
  number = {7383},
  pages = {63--67},
  publisher = {Nature Publishing Group},
  issn = {1476-4687},
  doi = {10.1038/nature10787},
  urldate = {2025-09-25},
  copyright = {2012 Springer Nature Limited},
  langid = {english}
}

@article{vijayanCavitymediatedLongrangeInteractions2024,
  title = {Cavity-Mediated Long-Range Interactions in Levitated Optomechanics},
  author = {Vijayan, Jayadev and Piotrowski, Johannes and {Gonzalez-Ballestero}, Carlos and Weber, Kevin and {Romero-Isart}, Oriol and Novotny, Lukas},
  year = {2024},
  month = may,
  journal = {Nature Physics},
  volume = {20},
  number = {5},
  pages = {859--864},
  publisher = {Nature Publishing Group},
  issn = {1745-2481},
  doi = {10.1038/s41567-024-02405-3},
  urldate = {2025-09-25},
  copyright = {2024 The Author(s)},
  langid = {english}
}

@article{ghorbani2025effects,
  title={Effects of quadratic optomechanical coupling on bipartite entanglements, mechanical ground-state cooling, and mechanical quadrature squeezing in an electro-optomechanical system},
  author={Ghorbani, N and Motazedifard, Ali and Naderi, MH},
  journal={Physical Review A},
  volume={111},
  number={1},
  pages={013524},
  year={2025},
  publisher={APS}
}

@article{jiang2016dynamics,
  title={Dynamics of an optomechanical system with quadratic coupling: Effect of first order correction to adiabatic elimination},
  author={Jiang, Cheng and Cui, Yuanshun and Chen, Guibin},
  journal={Scientific Reports},
  volume={6},
  number={1},
  pages={35583},
  year={2016},
  publisher={Nature Publishing Group UK London}
}

@book{burgess2020introduction,
  title={Introduction to effective field theory},
  author={Burgess, Cliff Peter},
  year={2020},
  publisher={Cambridge University Press}
}

@book{someda,
  title={Electromagnetic waves},
  author={Someda, Carlo G},
  year={2017},
  publisher={Crc Press}
}

@article{colemanRadiativeCorrectionsOrigin1973,
  title = {Radiative {{Corrections}} as the {{Origin}} of {{Spontaneous Symmetry Breaking}}},
  author = {Coleman, Sidney and Weinberg, Erick},
  year = 1973,
  month = mar,
  journal = {Phys. Rev. D},
  volume = {7},
  number = {6},
  pages = {1888--1910},
  publisher = {American Physical Society},
  doi = {10.1103/PhysRevD.7.1888},
  urldate = {2025-11-06}
}

@article{furutaniQuantumEffectiveAction2022d,
  title = {Quantum Effective Action for the Bosonic {{Josephson}} Junction},
  author = {Furutani, K. and Tempere, J. and Salasnich, L.},
  year = 2022,
  month = apr,
  journal = {Phys. Rev. B},
  volume = {105},
  number = {13},
  pages = {134510},
  publisher = {American Physical Society},
  doi = {10.1103/PhysRevB.105.134510},
  urldate = {2025-11-06}
}

@article{goldstoneBrokenSymmetries1962,
  title = {Broken {{Symmetries}}},
  author = {Goldstone, Jeffrey and Salam, Abdus and Weinberg, Steven},
  year = 1962,
  month = aug,
  journal = {Phys. Rev.},
  volume = {127},
  number = {3},
  pages = {965--970},
  publisher = {American Physical Society},
  doi = {10.1103/PhysRev.127.965},
  urldate = {2025-11-06}
}

@article{jona-lasinioRelativisticFieldTheories1964,
  title = {Relativistic Field Theories with Symmetry-Breaking Solutions},
  author = {{Jona-Lasinio}, G.},
  year = 1964,
  month = dec,
  journal = {Nuovo Cim},
  volume = {34},
  number = {6},
  pages = {1790--1795},
  issn = {1827-6121},
  doi = {10.1007/BF02750573},
  urldate = {2025-11-06},
  langid = {english}
}

\end{document}